# Opposites Attract – Muons as Direct Probes for Iodide Diffusion in Methyl Ammonium Lead Iodide

D. W. Ferdani,[1] A. L. Johnson,[1] S. E. Lewis,[1] P. J. Baker[2] and P. J. Cameron[1*]

[a]Centre for Sustainable Chemical Technologies, University of Bath, Claverton Down, Bath BA2 7AY, UK.

[b]ISIS Pulsed Neutron and Muon Source, Rutherford Appleton Laboratory, Harwell Oxford, Didcot OX11 0QX, UK.

*Correspondence and requests for materials should be addressed to PJC (email: P.J.Cameron@bath.ac.uk).

**The volume of research into organo-lead hailde perovskites is increasing rapidly, with perovskite solar cell efficiencies reaching as high as 22%. There is considerable evidence that mobile ions in the perovskite strongly influence the properties of the solar cell, with the majority of studies carried out on whole cells under bias. Here we use muon spin relaxation (µSR) to directly probe iodide diffusion in methyl ammonium lead iodide (MAPI). This is the first time that µSR has been used to detect iodide diffusion in any material and the results provide valuable insight into the movement of ions in lead halide perovskites. The experiment was carried out in the dark with no external biases applied and allowed us to calculate a diffusion coefficient of 1.6 x10-14 cm$^2$s$^{-1}$ for iodide in MAPI at 300 K.**

Ever since the first publication reporting perovskite solar cells (PSCs) appeared in 2009,[1] the technology has developed rapidly with efficiencies rising to 22.1 % in just 8 years.[2] Most research has focused on new materials[3–6] and processing methodologies[7–9] in a race to produce the highest efficiency solar cells. A clear understanding of the physical processes occurring within organo lead halide perovskites is still emerging.

Methylammonium lead iodide (MAPI) was the first perovskite to be widely investigated for solar cells and it is still used as a component of the most efficient PSC (e.g. Cs$_x$(MA$_{0.17}$FA$_{0.83}$)$_{100-x}$Pb(I$_{0.83}$Br$_{0.17}$)$_3$).[10–14] While the improvement in efficiency has been rapid the underlying physical properties of perovskite materials, such as molecular motion and diffusion, are still not fully understood. The consensus is that both electronic and ionic motion occur within perovskite materials.[15,16] Ion movement has been proposed as a cause of JV curve hysteresis in perovskite devices.[17–19] This finding is sometimes disputed as changes in solar cell contacts can reduce hysteresis in PSCs; however some careful experimental studies have demonstrated that JV curve hysteresis is linked to both ion movement and interfacial recombination.[20,21]

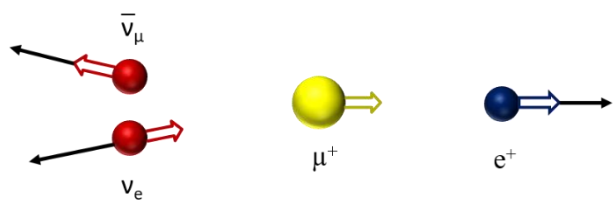

*Figure 1: Diagram depicting the decaying of a muon into a positron and two neutrinos.*

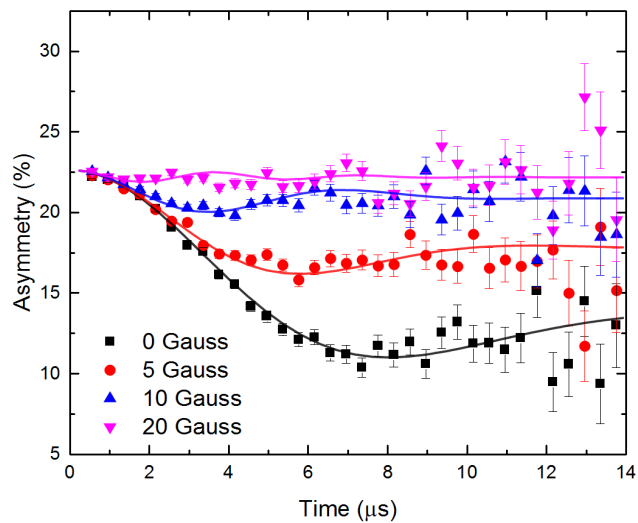

*Figure 2: Raw muon data for MAPI at 40 K with zero field (squares) and the applied longitudinal fields of 5 G (circles), 10 G (triangles), 20 G (inverted triangle). Also shown are the results of fitting the data to a dynamic Kubo-Toyabe function.*

There have been several computational and experimental studies on the movement of methyl ammonium ([MA]$^+$) and iodide ions in the perovskite material, producing a range of activation energies of 0.36 to 0.84 eV for [MA]$^+$ and 0.08 eV to 0.6 eV for iodide.[22–30] Theoretical first principles calculations performed by Eames et al. derived activation energies ($E_A$) of 0.58 eV and 0.84 eV for iodide and [MA]$^+$ diffusion respectively.[29] In their work, they also performed current-voltage response analysis which produced similar activation energies. They suggested that the diffusion path for iodide diffusion is slightly bowed around the edge of the lead iodide octahedra which is important for correctly calculating diffusion coefficients. Other calculations using different computational methodologies and predicted diffusion paths have found activation energies for iodide diffusion of 0.08 eV[22] and 0.44 eV.[28] Experimental measurements of iodide motion have also been carried and activation energies ranging from 0.17 eV measured by NMR[27] to 0.6 eV measured by chronoamperometry[29] have been reported. A range of other techniques have been used: an activation energy of 0.45 eV was found using capacitance measurements, 0.31 eV from temperature dependant current analysis, 0.5 eV from thermally stimulated current measurement and 0.55 eV and 0.43 eV using impedance spectroscopy.[23–27,31]

All of the previous methods, except for the NMR study, involve applying a voltage to and/or drawing a current from a complete PSC. When a voltage is applied to a complete device, there is evidence that iodide (and possibly also methylammonium) ions migrate to create ionic double layers at the interfaces. However, it can be difficult to tell the difference between changes due to ionic movement; changes due to degradation of the perovskite or changes in the contacts as a current is drawn. Some measurements have focused on studying ion migration by applying a large voltage across a thin section of a perovskite film, however at these high voltages degradation could also occur quite rapidly.[32] In this study, we have used muon spin relaxation studies to directly probe ion movement in MAPI crystallites. The material was measured in the dark and no current was drawn, the measurements allowed us to extract both an activation energy and a diffusion coefficient for the intrinsic movement of iodide inside the material.

The positive muons used in our experiment are unstable subatomic particles with similar properties to a positron, except their mass is 207 times larger and their lifetime is 2.197 μs. Muons have previously been used to study properties such as magnetism and superconductivity in a range of different materials.[33] Muon spin relaxation (μSR) has, more recently, been used to investigate the diffusion of Li⁺ and Na⁺ in modern battery materials.[34–38] Muons are implanted into the sample where they decay into a positron, which is most likely to be emitted in the muon spin direction at the instant of decay (Figure 1). The effect of local fields within the material on the muon spins is detected by the change in the asymmetry of the positron counts in detectors around the sample. Using this technique activation energies and diffusion coefficients of ions with nuclear magnetic moments can be measured. Previous μSR studies have focused on diffusion of lithium and sodium ions but the similarity between the nuclear magnetic dipole moment of iodide ions (+2.81 $\mu_N$) and lithium ions (+3.26 $\mu_N$) led us to investigate whether it would be possible to study iodide motion in MAPI using the same techniques that are applied to battery materials.[39] The ability of muons to act as discrete, non-destructive probes makes them ideal for studying the easily degraded perovskite material.

In this study μSR was carried out on powdered MAPI and per-deuterated $d_6$-MAPI. The un-deuterated and deuterated cations were compared so that it would be possible to observe if the cation had any effect on the muon spin relaxation data. We found that deuterating the [MA]⁺ had minimal effect on the crystal structure, as has previously been observed by Whitfield *et al*.[40] By comparing spectra taken with and without a $d_6$-[MA]⁺ cation we also hoped to be able to tell the difference between features caused by cation motion and anion motion. We first synthesised $d_6$-MAI with 99% of the hydrogens on the final product deuterated (Figure S1). $d_6$-MAPI was prepared using a hot casting method that is described in the ESI. Standard MAPI was also synthesised using the same hot casting method. Formation of the desired product was confirmed by pXRD analysis (Figure S2). The level of deuteration in the $d_6$-MAPI was measured using deuterium NMR (Figure S3) and found to be 83% in the final $d_6$-MAPI product. The loss of deuteration is likely to have been caused by the humidity of the processing environment; however, 83% deuteration should be sufficient to strongly affect any cation induced changes in the μSR.

μSR measurements were taken using a longitudinal field (LF) of 0, 5, 10 and 20 Gauss between 50 and 410 K with typical raw data shown in Figure 2. The fluctuation rate (ν) and the static width of the disordered local field distribution (coming from nuclear magnetic moments) at muon implantation sites (Δ) were subsequently calculated by fitting the muon asymmetry data at each temperature to a dynamic Kubo-Toyabe function multiplied by an exponential relaxation.[34] Four different LFs were used to decouple the muon spin relaxation from the local magnetic fields to different extents, allowing for more reliable fits to the data at each temperature.

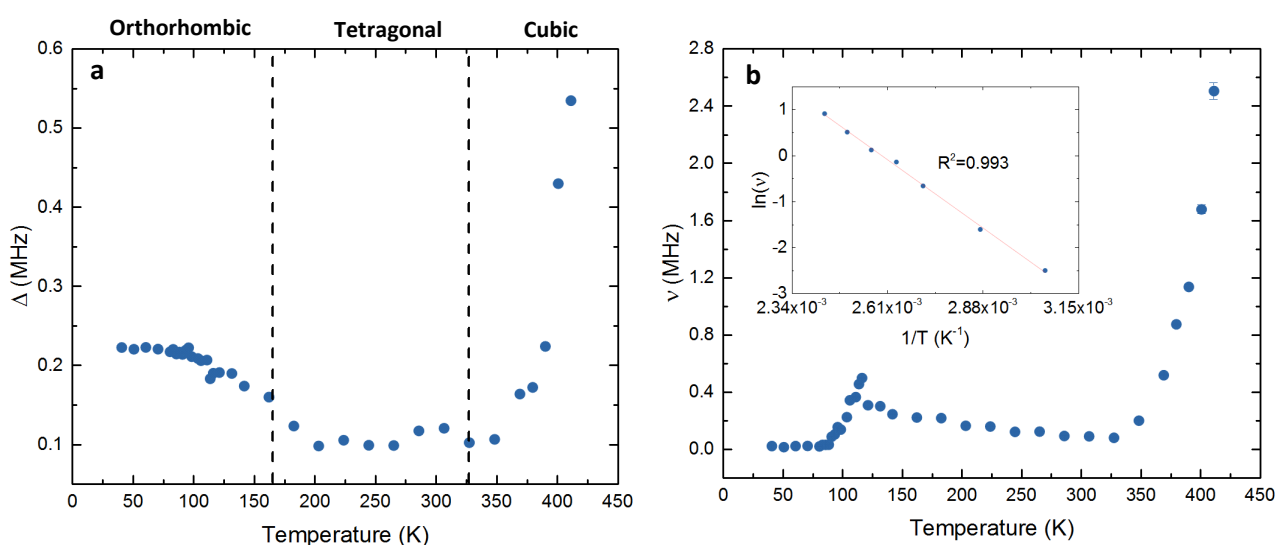

*Figure 3: Temperature dependence of (a) Δ and (b) ν values for MAPI derived from fitting raw μSR data to a Kubo-Toyabe function for measurements between 40 K and 410 K. The dashed lines in (a) indicate the phase transitions from orthorhombic to tetragonal and to cubic. The inset in (b) is the Arrhenius plot used to calculate the activation energy of the higher temperature process. Additional Arrhenius plots are in the supplementary information (Figure S4).*

Figure 3 shows the ν and Δ values plotted against temperature. The fluctuation rate ν initially plateaus between 50 K and 80 K before it begins to show a linear increase until 115 K. The increase in ν is typically caused by movement of ions in the material. The increase is followed by another plateau, before a large linear increase with an onset at 320 K. The activation energies ($E_A$) of these two processes were calculated by plotting an Arrhenius fit of these two linear regions (Figure 2b inset and Figure S4). The two activation energies calculated were 0.48 (±0.017) eV for the high temperature process and 0.073 (±0.004) eV for the low temperature. Δ initially decreases steadily from 50 K to 200 K before a plateau region followed by a large increase appearing at 350 K.

To probe whether the perturbations to the fluctuation rate, ν, were caused by the cation or anion, μSR was also run on per-deuterated $d_6$-MAPI. A comparison of the ν data for $d_6$-MAPI and MAPI is shown in Figure 4. The same two low temperature processes are evident for both samples, but the onset of the feature occurs at a slightly higher temperature in $d_6$-MAPI. The activation energies for the low temperature and higher temperature processes occurring in $d_6$-MAPI were calculated (Figure S5) and gave an $E_a$ of 0.098 (±0.008) eV for the low temperature process and 0.47 (±0.016) eV for the high temperature process. Deuteration of the [MA]$^+$ cation therefore leads to a ~ 10 K shift in the low temperature process, together with an increase in activation energy from 0.073 (±0.004) eV to 0.098 (±0.008) eV. As shown by Weller *et al.* and Whitfield *et al.*, at 100 K, MAPI and $d_6$-MAPI are in a less rigid orthorhombic structure with the methylammonium cations fully ordered. The {NH$_3$} groups align into the distorted square faces of the unit cell and the {PbI$_6$} octahedra are distorted.[40,41] Leguy *et al* used quasielastic neutron scattering and computational data to show that the [MA]$^+$ cations can reorient within the perovskite lattice at temperatures down to 140 K, temperatures below 140 K were not reported.[42] It seems likely that the low temperature response reported here is caused by the molecular motion of cations preceding the phase change at 165 K and it is interesting that this movement is seen so far below the phase transition temperature. As explained above, the onset for [MA]$^+$ movement occurs at a lower temperature than for d$_6$-[MA]$^+$. The onset of the high temperature feature occurs at the same temperature in both deuterated and non-deuterated samples, and the activation energies are nearly identical. This observation (along with the close agreement of the calculated activation and diffusion coefficients with literature values for iodide diffusion discussed below), leads us to believe that at higher temperatures we are directly detecting iodide diffusion in the perovskite.

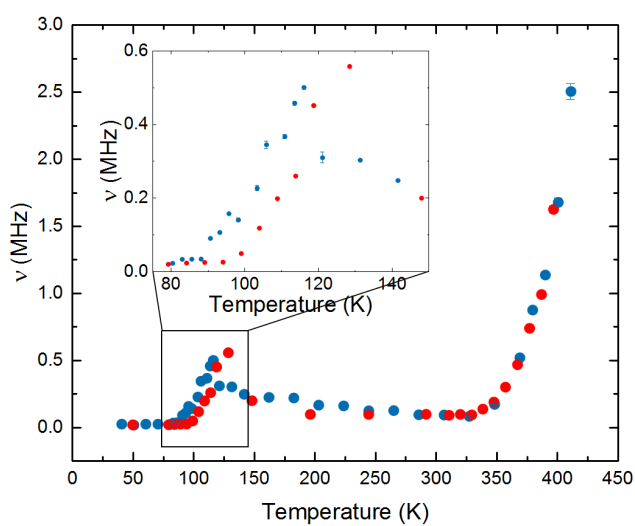

*Figure 4: Comparison of the ν of $d_6$-MAPI (Red) and MAPI (Blue) between 40 and 400 K with the inset being the same comparison but the temperature reduced to between 70 and 140 K to more easily see the difference in the low temperature process.*

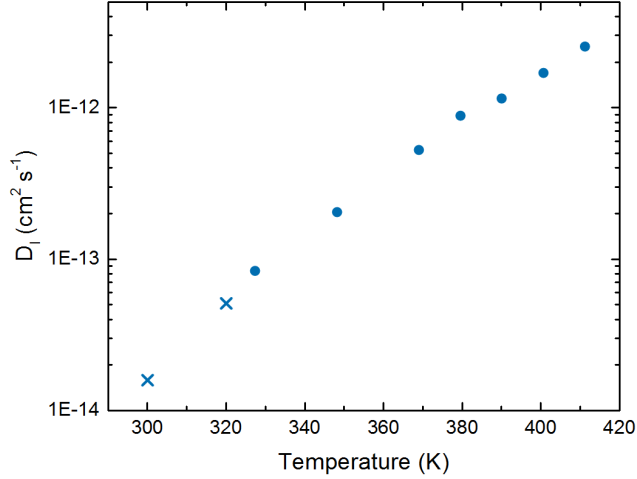

*Figure 5: A plot showing the change in $D_I$ between 300 and 410 K for MAPI with values calculated directly from the individually measured ν data points (circles) and values calculated by extrapolating ν data from the Arrhenius plot inset in Figure 2b (crosses).*

With the fluctuation data and Equation 1 it is possible to calculate the diffusion coefficient of iodide ($D_I$). Where $N_i$ is the number of accessible paths in the i-th path, $s_i$ the jump distance between iodide sites, $Z_{vi}$ the fraction of vacancies present and ν is taken from the muon data.

$$D_I = \sum_{i=1}^{N} \frac{1}{N_i} Z_{vi} s_i^2 \nu \qquad (1)$$

The diffusion coefficients (Figure 5) were calculated using the model for iodide diffusion presented by Eames et al.[29] e.g. by assuming there is just one mechanism for diffusion but eight possible pathways. A jump distance of 4.49 Å was used, calculated using neutron diffraction data for the inter atomic distances.[41] A vacancy fraction of 0.4%, as calculated by Walsh et al.[43] was also assumed. Extrapolating the linear section of the hopping data a diffusion coefficient of 1.6 x 10$^{-14}$ cm$^2$s$^{-1}$ was calculated at 300 K. Eames *et al.* predicted a diffusion coefficient of 10$^{-12}$ cm$^2$s$^{-1}$ at 320 K which is approximately two orders of magnitude greater than what we measure at 320 K (5.1 x 10$^{-14}$ cm$^2$s$^{-1}$).[29] The value of the calculated diffusion coefficient is sensitive to the vacancy fraction used. As can be seen from equation 1, doubling the vacancy fraction will double the diffusion coefficient at a given temperature. Diffusion coefficients of ~ 10$^{-14}$ cm$^2$s$^{-1}$ are at the lower end of what it is possible to measure using μSR and this may be the reason why iodide diffusion is only observed above room temperature as is discussed in more detail below.

During the μSR measurements the temperature of both perovskite samples was raised to 410 K. Perovskites are known to degrade at higher temperatures so it was important to check that the material had not been damaged by the measurement. In an attempt to unequivocally rule out degradation as a cause of the high temperature change in the fluctuation data, we investigated the stability of the sample in three different ways. Firstly, we performed thermogravimetric analysis on a

MAPI sample between 300 K and 450 K (Figure S6). No obvious mass loss was observed over this temperature range which agrees with previously observed thermal analysis on MAPI.[45] Secondly, pXRD analysis was performed on the MAPI sample both before and after the μSR experiments. Again, no change in the diffraction pattern was observed. The absence of peaks associated with the presence of $PbI_2$ strongly suggests no degradation of the perovskite material had occurred (Figure S6). Thirdly, after the initial muon measurements were taken, one of the samples was cooled to 100 K and several data points were repeated at a range of temperatures. There was no change in the data. We believe that the reason iodide diffusion is only observed above 320 K is because this is the temperature where the ionic hopping rate becomes large enough to be detected with μSR.[46] Diffusion is also happening at room temperature but we are not detecting it.

The performance and properties of PSCs made with MAPI and $d_6$-MAPI were also compared to ensure deuteration of the [MA]$^+$ had no significant effect of device performance. Average performance of both deuterated and non-deuterated cells was identical with no observable change in hysteresis (Figure S7).

As discussed above, the process observed above 320K is attributed to iodide diffusion in the perovskite. Table 1 compares our calculated activation energy with others presented in the literature, both from computational studies and a range of different experimental techniques.

The activation energy measured in this study (0.48 eV) is remarkably close to values obtained from a range of electrochemical measurements. It has been suggested that the processing method used to prepare the perovskite influences the activation energy for ion movement.[24] Small differences in activation energy between samples could be due to changes in crystallinity and defect density resulting from the wide range of methods used to prepare perovskite thin films and powders.[47] The benefit of muon spectroscopy is that it measures the intrinsic ion diffusion in the material without the influence of interfaces.

The Δ for both powders show a similar trend with a gentle decrease before 200 K, followed by a plateau, before a dramatic increase starting at 350 K. The steady decrease in Δ between 80 and 120 K is likely to be caused by motional narrowing as the movement of ions causes a decrease in the static field distributions. The large decrease after 150 K can be attributed to a change in the structure of the perovskite from orthorhombic to tetragonal which occurs at 165 K where the plateau in Δ values begins. The final increase in Δ occurs after the phase transition from tetragonal to cubic at 327 K. It is likely that the change in structure changes the nuclear magnetic fields at the muon stopping site that are described by Δ and we do not reach a high enough temperature to reach a new plateau.

*Table 1: Comparison of literature values for the activation energy of iodide diffusion.*

| Method | $E_a$ (eV) Computational | $E_a$ (eV) Experimental |
|---|---|---|
| Thermally Stimulated Current Measurement[32] | - | 0.5 |
| Capacitance[25] | - | 0.45 |
| Temperature Dependant Current[24] | - | 0.31 |
| Intensity Modulated Voltage Spectroscopy[23] | - | 0.55 |
| Impedance[26] | - | 0.43 |
| Temperature Dependant Impedance[44] | | 0.58 |
| NMR[27] | - | 0.17 |
| Temperature Dependant JV Curves[48] | | 0.33 |
| First Principles & Chronoamperometry[29] | 0.58 | 0.6 |
| First Principles[28] | 0.44 | - |
| First Principles[31] | 0.08 | - |
| **This study** | - | **0.48** (±0.017) |

## Conclusions

In summary, we report the use of μSR to detect iodide diffusion for the first time. We show that iodide moves in MAPI close to room temperature. The activation energy was calculated to be 0.48 eV and the diffusion coefficient was $1.6 \times 10^{-14}$ cm$^2$s$^{-1}$ at 300 K. The diffusion coefficient is two orders of magnitude lower than the computationally predicted value. By using a perdeuterated analogue of MAPI we confirmed that the higher temperature process is due to iodide motion and is not related to cation movement. μSR is an ideal technique to study iodide diffusion in MAPI and no degradation of the material occurs during the measurement.

## Conflicts of interest

There are no conflicts to declare.

## Acknowledgements

DWF acknowledge funding from the EPSRC Centre for Doctoral Training in Sustainable Chemical Technologies: EP/G03768X/1. The authors would also like to thank ISIS Pulsed Neutron and Muon Source for allocation of experimental beamtime on EMU under project RB1700011. PJC and PJB acknowledge Prof. Saiful Islam for suggesting they talk to each other about using μSR to measure lead halide perovskites.

## Notes and references